\documentclass[final,twocolumn,superscriptaddress,aps,prb,a4]{revtex4} 
\usepackage{latexsym}
\usepackage{times} 
\usepackage{graphicx}
\usepackage[tight,FIGTOPCAP]{subfigure}

\begin{document} 

\title{Hierarchy and Feedback in the Evolution of the {\em E.~coli}
  Transcription Network.}

\author{M.~Cosentino Lagomarsino} 
\altaffiliation{and Universit\`a degli Studi di Milano, Dip.
    Fisica.  Via Celoria 16, 20133 Milano, Italy} 
\email[ e-mail address: ]{mcl@curie.fr}
\affiliation{UMR 168 / Institut Curie, 26 rue d'Ulm 75005 Paris, France}
\author{P.~Jona} 
\affiliation{Politecnico di Milano, Dip. Fisica, Pza Leonardo Da Vinci
  32, 20133 Milano, Italy} 
\author{B.~Bassetti}  
\altaffiliation{and I.N.F.N. Milano, Italy. Tel. +39 - (0)2 - 50317477
  ; fax +39 - (0)2 - 50317480 } 
\email[e-mail address: ]{ bassetti@mi.infn.it } 
\affiliation{Universit\`a degli Studi di Milano,
  Dip.  Fisica. Via Celoria 16, 20133 Milano, Italy}

\author{H.~Isambert} 
\email[ e-mail address: ]{herve.isambert@curie.fr}
\affiliation{UMR 168 / Institut Curie, 26 rue d'Ulm 75005 Paris, France}

\begin{abstract}
  The {\em E.coli} transcription network has an essentially
  feedforward structure, with, however, abundant feedback at the level
  of self-regulations. Here, we investigate how these properties
  emerged during evolution. An assessment of the role of gene
  duplication based on protein domain architecture shows that (i)
  transcriptional autoregulators have mostly arisen through
  duplication, while (ii) the expected feedback loops stemming from
  their initial cross-regulation are strongly selected against. This
  requires a divergent coevolution of the transcription factor
  DNA-binding sites and their respective DNA cis-regulatory regions.
  Moreover, we find that the network tends to grow by expansion of the
  existing hierarchical layers of computation, rather than by addition
  of new layers. We also argue that rewiring of regulatory links due
  to mutation/selection of novel transcription factor/DNA binding
  interactions appears not to significantly affect the network global
  hierarchy, and that horizontally transferred genes are mainly added
  at the bottom, as new target nodes.  These findings highlight the
  important evolutionary roles of both duplication and selective
  deletion of crosstalks between autoregulators in the emergence of
  the hierarchical transcription network of {\em E.coli}.
\end{abstract}

\maketitle

The successful adaptation of microorganisms to an environment or host
is determined by the correct response to external and internal stimuli
through the simultaneous expression of a large set of genes.
The basal mechanism that performs this task is transcriptional regulation, so
that it becomes important to characterize this regulatory process from a
global, or ``network'' viewpoint.
Transcriptional regulation networks are defined starting from the basic
functional elements of transcription~\cite{BLA+04}.  To construct the
associated graph, one usually represents each operon with a node, and each
regulatory interaction with a directed link $A \rightarrow B$ between the
target operon $B$ and the operon $A$ coding for a transcription factor (TF)
that has at least one binding site in the cis-regulatory region of $B$. 
A transcription factor regulating its own expression is
  called an autoregulator (AR).
With this definition, the interaction graph structure is accessible by
large-scale and collections of small-scale
experiments~\cite{SMM+02,SSG+06,LRR+02,HGL+04}.

Some topological and evolutionary properties of transcription networks
have been elucidated~\cite{MIK+04,WtW04,TB04}.  In particular, they
can be analyzed in terms of a hierarchy of inputs that produce output
responses~\cite{MBZ04,MKD+04,YG06}.
Specifically, the \emph{ E.~coli} transcription network has an essentially
feedforward layered structure, where feedback is mainly limited to
autoregulations~\cite{MBZ04,MKD+04}. The abundance of the latter is, however,
striking, as they concern more than half of the transcription
factors~\cite{THP+98}.
Here, after quantifying the marginality of these properties with respect to a
null network ensemble, we investigate how they could have emerged during
evolution.  An assessment of the role of gene
duplication based on protein domain architecture
  shows that {\em i)} transcriptional autoregulators have mostly arisen
  through duplication, while {\em ii)} the expected feedback loops stemming
  from their initial cross-regulation.
  are strongly selected against.  This requires a {\em divergent coevolution}
  of the autoregulator DNA binding sites and their respective DNA
  cis-regulatory regions.  Moreover, we find that the network shows a tendency
  to grow by expansion of the existing hierarchical layers of computation,
  rather than by addition of new layers.
  We also argue that {\em de novo} rewiring of regulatory links due to
  mutation/selection of novel transcription factor/DNA binding interactions
  does not affect the hierarchy, and that horizontally transferred genes are
  mainly added at the bottom, as new target nodes.  
  Our findings are consistent with a view of prokaryote evolution based on
  ancient duplications and conservation of stable central parts despite
  widespread horizontal gene transfers~\cite{CD04,PPL05}.

\paragraph{Feedback and Hierarchy.}

\begin{figure}[htbh]
  \centering \includegraphics[width=0.47\textwidth]{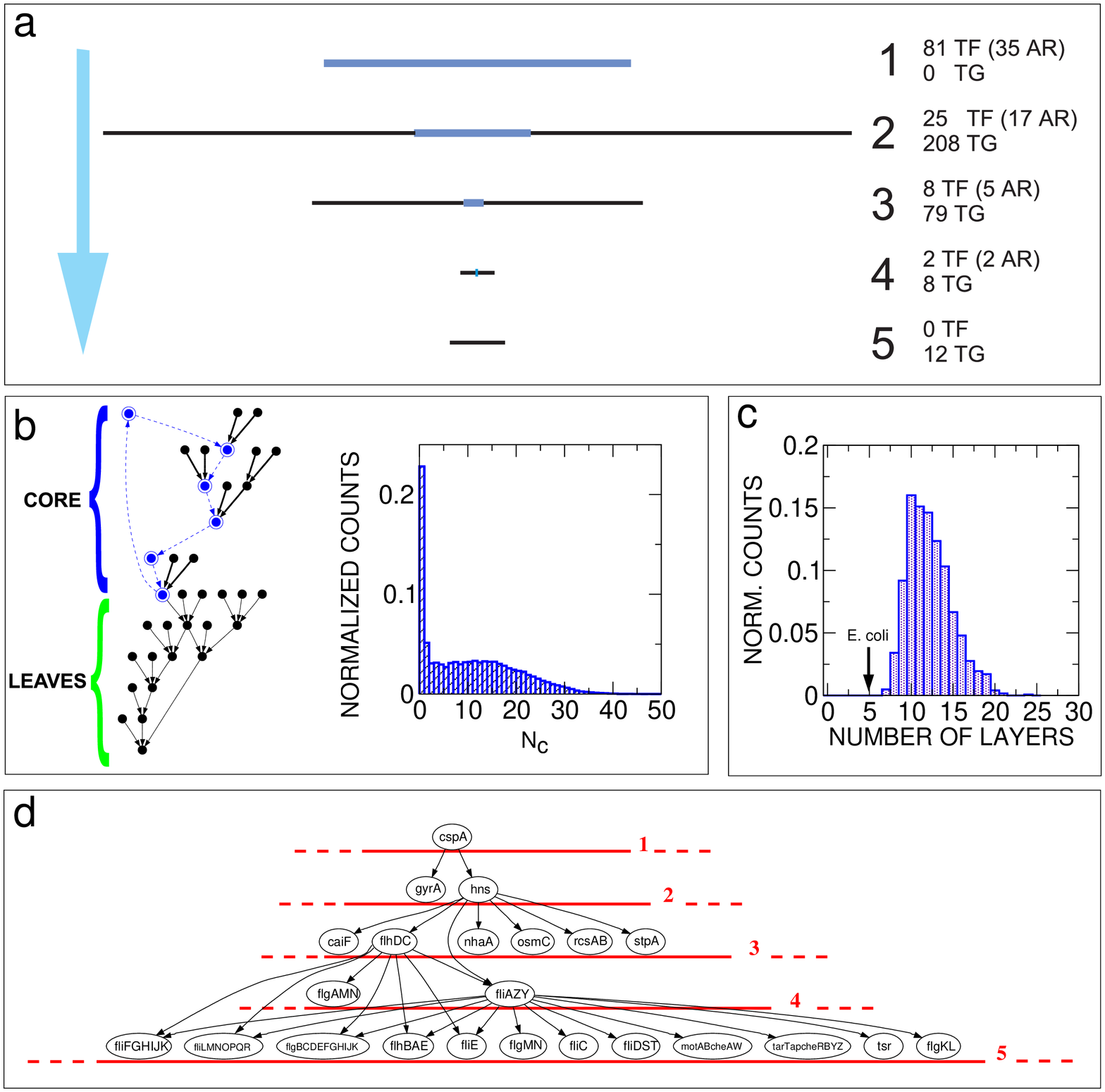}
  \caption{{\small Feedback and hierarchy in {\em E.~coli}
    transcription network.  (a) Scheme of the layer structure of the
    network. Direction of regulatory links is from top to bottom. Each
    line represents a layer, populated by TFs (blue, thick line) and
    TGs (black, thin line). Members of layer $i$ are regulated at most
    by $i-1$ nodes plus themselves. By definition, layer one is
    constituted entirely by TFs. Annotations on the right hand side of
    the layers specify their population of TGs, TFs and ARs.
    (b) Evaluation of feedback with the leaf-removal algorithm. Right:
    illustration of the leaf-removal algorithm. Leaves are nodes that do not
    regulate any other node. Removal of one leaf and its regulatory links may
    create a new leaf. Iterative removal of leaves has to stop at a core of
    nodes that contains loops (blue, circled nodes, dashed links).  The core
    might contain tree-like components upstream of the loops (black). Left:
    histogram of the number of nodes in the core $N_C$ for randomized
    counterparts of {\em E.~coli}~\cite{LJB05}.
    The data refer to $1.1\cdot10^{6}$ accepted MCMC moves for randomization
    (see methods and Supplementary Note S1).
    (c) Histogram of the layer number in the randomized counterparts of the
    {\em E.~coli} network. The average number of observed layers is
    about $12$, to   
    compare to the $5$ of {\em E.~coli}. The data correspond to a MCMC
    run where a 
    total of $5.78\cdot10^{8}$ matrices where generated (of which about
    $1.23\cdot10^{8}$ were tree-like).
    (d) The flagella-building subnetwork is the only example of
    functional subnetwork that spans all the five layers. Here, this
    subnetwork is constructed arbitrarily starting from a member of
    layer one and following the tree downstream.}}
  \label{fig:fback}
\end{figure}

{\em A priori}, one may expect that transcription networks contain
abundant feedback loops involving two or more
genes~\cite{Tho73,LJB05}. However, for the case of {\em E.~coli}, the
available data indicate that this is not the
case~\cite{MBZ04,MKD+04,YG06}.
The Shen-Orr dataset~\cite{SMM+02} (423 operons; 117 TFs, 578
interactions) does not contain any non-self-regulatory feedback loop
for the {\em E.~coli} transcription network.  Such a tree-like
directed graph is naturally organized in feedforward layers of
computation, ending with target genes (TG) as ``leaves''.
The layers and their numbering can be defined by the longest chain of
(different) regulators upstream of each TF or TG in each layer
(Figs.~\ref{fig:fback}a\&d).  Members of layer one are regulated by at
most themselves, members of layer two are regulated by a chain of one
transcription factor and possibly themselves, and so on.
There are five hierarchical layers in the Shen-Orr
dataset~\cite{SMM+02}, which is considerably lower than for randomized
null networks (see Fig.~\ref{fig:fback}c). About $50\%$ of the nodes
(TFs and TGs) lay in layer two, with $69\%$ of all TF nodes located in
layer one.
The notable exception to this general lack of feedback is the
substantial presence of feedback loops involving a single node, or
autoregulators (59 ARs out of 117
TFs)~\cite{WHS04,BS00,ASM+03,THP+98}.
The more recent publicly available database
RegulonDB~5.5~\cite{SSG+06} includes larger
datasets~\cite{MBZ04,MKD+04,SSG+06} (648 operons; 147 TFs, 1170
interactions, 85 ARs, excluding Sigma-factor interactions). By
contrast with Shen-Orr dataset, it contains a few (4)
non-self-regulatory feedback loops and a few more (a total of 7)
hierarchical layers but still considerably less than in randomized
null networks (see Supplementary Note S5).  Hence the same trend is
seen for both Shen-Orr and RegulonDB~5.5~\cite{SSG+06} datasets.
%
%

To quantify the significance of regulatory feedback and hierarchical
properties of the {\em E.~coli} transcription network, we compared it
for each dataset (Shen-Orr and RegulonDB~5.5) with randomized null
networks with the same degree sequence, {\em i.e.}  conserving the
number of incoming and outgoing links for each node
(Fig.~\ref{fig:fback} and Supplementary Note S1).
For both data sets, the number of ARs found in the empirical network
greatly exceeds the same quantity for randomized counterparts,
confirming previous observations on self-regulatory
feedback~\cite{ASM+03,THP+98,SMM+02}.
The importance of non-self-regulatory feedback was
quantified by the size of the regulatory core obtained after pruning
the tree-like input and output cascades using the leaf-removal
algorithm (see Fig.~\ref{fig:fback}b and Supplementary Notes S1 and
S5).  From this analysis, we conclude that the importance of
transcriptional, non-self-regulatory feedback is significantly lower
in both empirical networks (Shen-Orr and RegulonDB~5.5) than in their
randomized network counterparts, see Fig.~1b and Fig.~S5.10.

The importance of hierarchy was also quantified.  As there is no
straightforward definition of hierarchy in general for networks
including feedback, we have used the total number of layers in the
tree-like input and output branches of the network as practical
definition of hierarchy.  This also corresponds to the number of
iterations of the leaf-removal algorithm (see, however, alternative
definitions of hierarchy in Supplementary Note S5).  Note, in
particular, that it correctly recovers the actual number of
hierarchical layers for tree-like directed graphs (overlooking
possible self-regulatory links as in the case of Shen-Orr dataset).
Comparisons with null models were restricted to randomized networks
with the same regulatory core size.  Remarkably, the number of
hierarchical layers was found to be considerably lower than in typical
randomized network counterparts for both Shen-Orr and RegulonDB~5.5
datasets, see Fig.~1c and Fig.~S5.11 and Supplementary Note S5.

\paragraph{Evolutionary Drives.}

What is the evolutionary origin of this peculiar structure? There are
three main mechanisms for the evolution of a transcription network.
(1) Gene duplication, (2) rewiring of links by mutation/selection of
TF/DNA interactions (3) horizontal gene transfer.  All three
mechanisms, which we discuss below in the context of transcription
network evolution, have been shown to play a substantial role in
prokaryote evolution~\cite{BLA+04,PPL05,TB04,CW03,DMA05,MBV05}.  For
clarity, the following discussion refers only to the Shen-Orr dataset,
which is still to date the most widely used dataset.  The same
detailed analysis on the RegulonDB 5.5 dataset is discussed in a
dedicated section S5 in the Supplementary Note.

\subparagraph{Duplication.}

Following previous analyses~\cite{TB04,MBT03},
we define proteins that are likely to share a common ancestor through
structural domain assignments of the SUPERFAMILY database~\cite{GKH+01}.
These domains allow for the definition of larger classes than sequence
comparison alone~\cite{TB04}.  The database enables to associate an ordered
sequence of domains, or ``domain architecture'' to each protein.
We define protein homologs as proteins whose domain architectures are
identical neglecting domain repeats~\footnote{This corresponds to a
  conservative view of homology where no domains are acquired or lost after
  duplication.  More flexible and realistic definitions of homologs, yield
  essentially the same results (Supplementary Note S2)}.
We have analyzed the distribution of regulatory links between and within
classes of likely duplicate genes.  The statistical significance of the
analysis in terms of homology classes is established~\cite{TB04} by comparison
with random shufflings of genes (TFs and TGs separately) between classes.
\begin{figure}[bth]
  \centering
    \includegraphics[width=0.45\textwidth]{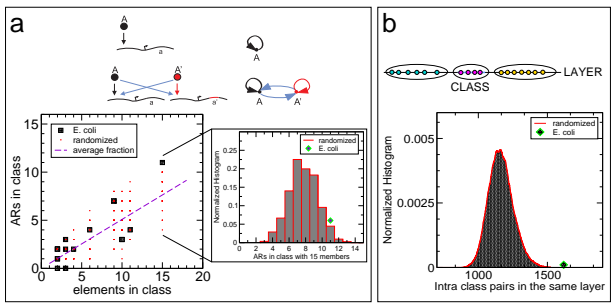}
  \caption{{\small Duplication of ARs in the {\em E.~coli}
  transcription network.
    (a) 
    ARs are propagated by duplication of the network (See also Table 1a), and
    need to develop specificity by coevolution.
    Top: the mechanism for duplication. A is an AR. In an initial stage,
    the original, A, and its copy, A', are identical. This creates a circuit
    where both A and A' are ARs, and there is mutual crosstalk (light blue)
    links. Subsequent divergence can erase the links (See Supplementary Note
    S3).
    Bottom left: population of ARs in the homology classes in the {\em E.~coli}
    network with original vs randomized domain associations. The $x$ axis
    reports the size of each class of transcription factors, while the $y$
    axis indicates the fraction of autoregulators in the class. The dashed
    line corresponds to the expected value computed from the total fraction of
    ARs. Red dots are randomized instances. 
    Bottom right: Histogram of the AR population (number of ARs in the class)
    of the largest homology class (having 15 members) for $10^{5}$
    randomizations of the SUPERFAMILY structural domains of the TFs, compared
    to the observed quantity in {\em E.~coli} (diamond). In most ($95\%$) of
    the randomizations the class with 15 members contains less than 11 ARs,
    indicating that duplication is likely.
    (b) Layers tend to be populated by members of the same homology class.
    Comparison with randomizations of the structural domain associations of
    all the genes. The $x$ axis reports the total number of gene pairs of the
    same homology class belonging to the same layer. The histogram represents
    the randomized case, while the diamond indicates the observed value in
    {\em E.~coli}.}}
  \label{fig:ardupl}
\end{figure}

The first result, summarized in Table~\ref{tab:tables}a (see also
Supplementary Table~S5.1a), shows that duplicates of ARs tend to
retain their self-links.  We quantified this using two global
parameters, $h_{ar}$ and $g_{ar}$. $h_{ar}$ is the average fraction of
ARs in classes with two or more ARs. It measures the tendency to have
many ARs in one class if two are already present (the reason of the
cutoff is to exclude from the count classes with two members and only
one AR).  $g_{AR}$ is the variance across classes of the fraction of
ARs within a class. This parameter measures the tendency to have
classes that are more populated than average, and at the same time
classes that are less populated than average, which can be observed in
Fig.~\ref{fig:ardupl}a (and Supplementary Fig. S2.5).
In spite of this strong evidence for the proliferation of ARs through
duplication events, we already mentioned the absence of {\em any} two-node
feedback loops between homologous (or non-homologous) ARs~\footnote{This is
  not strictly true for the more recent RegulonDB 5.5 dataset, where a few of
  these two-node feedback loops are observable, though the signature for
  negative selection remains (see Supplementary Note S5)}.  This requires that
the initial cross-regulation between duplicated ARs (reflecting the fact that
binding sites are initially identical) is systematically suppressed even if
self-regulation is conserved for both TF copies (Fig.~\ref{fig:ardupl}a). We
also find that \emph{single} regulatory links between any kind of TFs in the
same homology class are very scarce and always involve at least one AR (see
Fig.~S2.7).  On average, $91\%$ of the links within a homology class of TFs
are self-links.

A simple duplication-divergence model (Fig.~\ref{fig:ardupl}a and
Supplementary Note S3) shows that the {\em concomitant} conservation of
self-links and cancellation of cross-talks between duplicated ARs require a
selective pressure for evolutionary decoupling.
This can be achieved through \emph{divergent coevolution}~\cite{PKT06,TB04} of
duplicate TF/DNA binding interactions.  For instance, a straightforward
analysis of the binding sites of CRP and FNR, two duplicate ARs regulating
many TGs having no cross-regulation, shows that their own DNA cis-regulatory
regions have higher specificities than the cis-regulatory regions of most of
their TGs (See Supplementary Note S4), which suggests decoupling of their
self-regulatory links.

\subparagraph{Layer Hierarchy and Rewiring.}

As shown in~\cite{TB04}, a large fraction of the non-self-regulatory links of
the {\em E. coli} transcription network likely originated from duplication
events.  Indeed, many pairs of TGs from the same homology class are regulated
by a common TF; likewise, many homologous TFs regulate the same TGs, and many
pairs of TFs from the same homology class regulate homologous pairs of TGs.
Clearly, the likely duplication events underlying this transcription network
expansion conserve the number of TFs upstream of each target, hence leaving
the layer hierarchy untouched.  The only duplication event that can actually
add a layer is the duplication of an AR, provided that a crosstalk is
conserved.  A comparison of the homology classes with the populations of the
network layers (Fig.~\ref{fig:ardupl}b, Table~\ref{tab:tables}b, Supplementary
Table~S5.1a, and Supplementary Note~S2), shows that globally genes of the same
homology class tend to populate the same layer.

In fact, we find only 5 non-self-regulatory links within homology classes (see
Supplementary Fig.~S2.7) and they all involve at least one AR, suggesting that
they originated from duplication events of an AR.  For example, the
histone-like autoregulator H-NS, belonging to layer 2, regulates its homolog
StpA, which belongs to layer 3 (Supplementary Fig.~S2.7).
Yet, the coincidence between the number of non-self-regulatory links within
homology classes and the number of hierarchical layers in {\em E. coli}, does
not allow to conclude that the layers were generated by AR duplication events.
Evidence for some presumed rewiring of regulatory links also exists. For
instance, the same AR H-NS (Supplementary Fig.~S2.7) is also regulated by the
cold shock protein CspA, which neither regulates any homologs of H-NS, nor has
any homolog itself in the dataset.  It is thus likely that this incoming
regulatory link of H-NS does not come from duplication, but rather, from
rewiring. Thus rewiring could also be a mechanism for creation of new
computational layers.
However, we find also indications that {\em de novo} rewiring of regulatory
links is limited by the network hierarchy.
With respect to randomized instances, there is smaller dispersion of TG
homology classes over multiple layers of computation than observed for TF
homology classes. This can be quantified for example by the Z-score of the
number of gene pairs in the same layer and class; the higher this quantity,
the more duplication dominates on rewiring. We find $Z=1$ for layer one
(entirely made of TFs), while $Z=4.6$ for layer two (dominated by TGs).
This is consistent with an evolutionary scenario leading to an early
structuration of the transcriptional network into a few hierarchical layers of
computation (from duplication of ARs and limited rewiring as well) followed by
a primarily lateral expansion of TGs (mostly by duplication).

Altogether, these observations lead us to conclude that maintaining a
``shallow'' layer structure, where most of the computation is
performed at the single layer level, seems to be important for the
{\em E.~coli} transcription network. A possible rationale for this
fact is that the time taken by a computational cascade involving
multiple layers is expected to be roughly proportional to the number
of layers~\cite{REA02}. Thus, since the network has to react to a
particular stimulus or environment by ``switching on'' the proper
genes without unnecessary delays, having many layers might be
disadvantageous.  For this reason, it could be interesting to target
studies to the sub-systems that make use of multi-layer computation
(Fig.~\ref{fig:fback}d).

\begin{table}[tbh]
  \centering
  \caption{{\small Evaluation of different evolutionary drives (see
    also Supplementary Table~S5.1).
(a) The table shows that duplicates of ARs tend to retain their
self-links. This is  quantified globally by the observables $h_{ar}$, the
average   fraction of ARs in classes with two or more ARs, and $g_{ar}$,
measuring the spread in  the AR population among classes that can be observed
in Fig.~\ref{fig:ardupl}a and Supplementary Fig.~S2.5.
(b) Duplication and divergence preserve the layer structure. The first column
indicates distance between layers (defined as the absolute difference in layer
numbers), while the second and the third correspond to
the population of duplicate genes (genes in the same homology class) at that
distance, in $10^{5}$ instances with randomized domain associations (average
values) and the E~.coli  domain association dataset respectively. For example,
the first row (pairs of genes at distance zero) concerns  the number of
duplicate genes which occupy the same layer (see Fig.~\ref{fig:ardupl}b and
Supplementary Note S2). The sketch in the right panel illustrates the
distribution of nodes belonging to the same class of TFs (cyan) or TGs
(yellow) among the layers, and the definition of distance between layers.  
(c) Fate of gene gains from horizontal transfer. TFs are underrepresented both
in the class of gene gains (columns 2 and 3) and in the class of gene gains
that have at least a paralog in the homology classes constructed with domain
associations (columns 5 and 6).    
}}
  \includegraphics[width=0.45\textwidth]{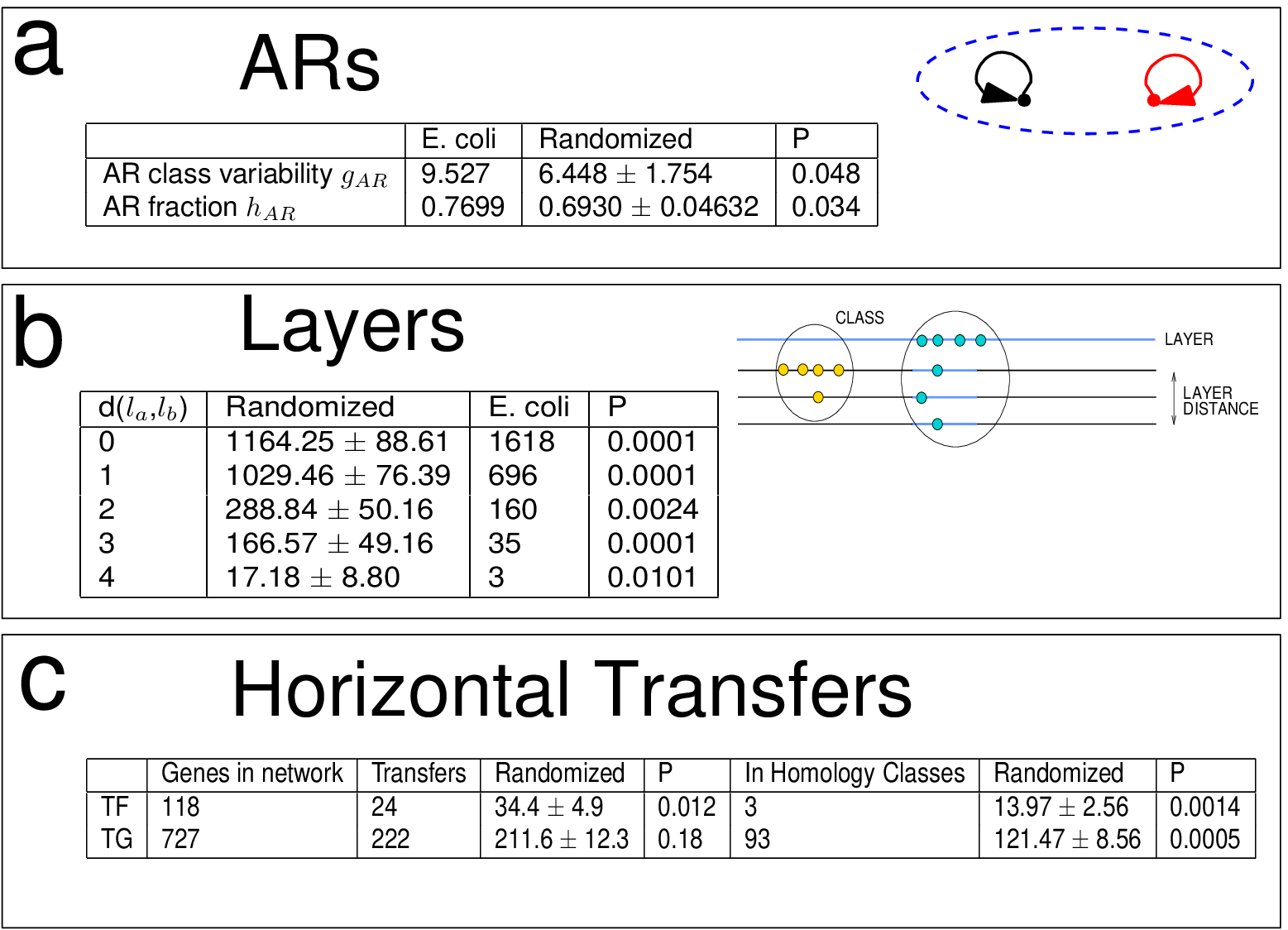}
  \label{tab:tables}
\end{table}

\subparagraph{Horizontal Gene Transfer.}  
Finally, let us focus on horizontal gene transfer.  We investigated the
role of transferred genes with respect to their position in the network and in
the homology classes (Table~\ref{tab:tables}c, Supplementary Table~S5.1b).
For this purpose, we used lists of genes likely to be transferred in {\em
  E.~coli} from ref.~\cite{PPL05}.  These lists were obtained by a
phylogenetic tree reconstruction based on 51 bacterial species.  With a
gain/loss penalty of two, $29\%$ of the genes in the network are classified as
gene gains.
We find that most of the gene gains are target genes. Comparison with a simple
binomial null model shows that most TFs are not likely to have been
horizontally transferred, while transferred TGs are abundant.  Hence, one can
conclude that in analogy with {\em E.~coli} metabolic network~\cite{PPL05},
imported genes are mainly found at the ``periphery'' of the network.
Furthermore, transferred TGs are not found in large homology classes,
defining instead mostly single-gene classes, suggesting that gene
duplications preceded many horizontal gene transfers.  Overall, this
is consistent with a view of prokaryote evolution based on ancient
duplications and conservation of a ``stable genetic core'' despite
widespread horizontal gene transfers~\cite{CD04,PPL05}.

\vskip 0.5cm 


In conclusion, our findings confirm the importance of (probably
ancient) duplications for the evolution of this network, and pinpoint
to some important trends due to selective pressure and evolutionary
dynamics, namely, preservation of ARs and cancellation of crosstalks,
as well as a propensity for a feedforward structure with a small
number of computational layers. The layered hierarchy of {\em E. coli}
transcription network appears to have first emerged and laterally
expanded from duplication of a few ARs.  Overall, this supports an
evolutionary scenario based on duplication~\cite{TB04} (with
duplicates occupying the same layer) and selective deletion of
crosstalks between autoregulators (which would otherwise increase the
number of hierarchical layers).  Further duplication-driven lateral
expansions of TG homology classes have then taken place together with
widespread horizontal gene transfers of new TGs.

\section*{METHODS}
\label{sec:methods}

\paragraph{Datasets.}
We used the Shen-Orr and RegulonDB~5.5 datasets for the transcription
network~\cite{SMM+02,SSG+06}. Domain architecture data were taken from
the SUPERFAMILY database~\cite{GKH+01}, version 1.61, as in the
datasets of ref.~\cite{TB04}.
More recent versions of SUPERFAMILY (we tested 1.69) or the transcription
network~\cite{SSG+06} do not change the conclusions.
The dataset of likely horizontally transferred genes was generously provided
by the authors of ref.~\cite{PPL05}.
Finally, the binding sites for the clustering analysis FNR and CRP (see
Supplementary Note S4) were taken from the regulonDB~\cite{SSG+06} dataset.

\paragraph{Network Analysis.}
We used Fortran 77 implementations of different variants (see Supplementary
Note S1) of the leaf-removal algorithms on the Shen-orr data-set (including
ARs) and its randomized counterparts, which were obtained using a standard
Markov Chain Monte Carlo (MCMC) algorithm that preserves the degree sequence
(marginals of the adjacency matrix)~\cite{RJB96}.  This algorithm is best
formulated for the adjacency matrix of the graph, i.e.  the matrix $A$ such as
$(A)_{ij} = 1$ if $i \rightarrow j$, and 0 otherwise. We considered
\emph{unstructured} counterparts of $A$. Randomizations with no self-links or
structurally zero diagonal of $A$, lead to different results.  For all the
tree-like instances, the number of layers correspond to the (whole-graph)
iterations that are necessary for the leaf-removal algorithm to remove the
entire graph. In order to consider a significant sample, the number of MCMC
iterations was calibrated according to the number of accepted MCMC
moves~\cite{RJB96}. Specifically, we stopped the algorithm after $T=K \tau$
accepted moves, where $\tau$ is the number of nonzero elements of $A$, and $T=
2000$.

\paragraph{Evaluation of Duplications.}
We constructed classes of homologous genes using similarity criteria of the
SUPERFAMILY domain architecture.  Results given in the body of the paper refer
to the case where two genes are considered homologs if they share the same
domains in the same order, neglecting domain repeats. A gap is considered a
domain. Different choices lead to very similar results (see Supplementary Note
S2).  For this analysis, proteins coded by the same operon were considered as
separate entities. Many classes generated this way, such as $\{CRP,FNR\}$, are
supported by evidence based on protein sequence comparison.  The classes of
proteins obtained this way were compared with TF-TG links in the transcription
network data-set.  Observations related to these classes were compared to
randomizations that shuffle domain associations to gene names, separately for
TFs and TGs~\cite{TB04}. The data given in the body of the paper correspond to
$10^5$ randomizations.

\paragraph{Graph Growth Model.}
A simple model of duplication-divergence was considered, where at each time
step duplication of the graph is followed by cancellation of links with
prescribed probabilities (Supplementary Note S3).  
We analyzed the evolution equations for the fraction of ARs and of intra-class
links, in the different scenarios of symmetric and asymmetric divergence,
presence or absence of selective conservation of ARs, presence or absence of
constant inflow of ARs. The results were compared with the observed trends in
the data.

\paragraph{Analysis of Horizontal Gene Transfers.}

We used lists of imported genes obtained by a phylogenetic tree reconstruction
based on 51 bacterial species~\cite{PPL05}.  We presented results obtained
with a gain/loss penalty of two and the hypothesis of retarded transfer, or
``DELTRANS'' assumption. Different choices lead to similar results (data not
shown).
To evaluate the partition of transferred genes between TFs and TGs, we
compared with a simple binomial model where the probability of import is given
by the total fraction of imported genes.  As a null model for the number of
imported genes that appear in homology classes, we considered classes
generated by shuffling associations of genes with domain architectures as
above.

\paragraph{Specificity of TF Binding Sites.}
Binding sites of two duplicate TFs were scored against their
logos~\cite{Sch02}, obtained with the list of all available binding sites from
RegulonDB.  The specificity was defined as the difference between the scores
of the same binding sites on two different logos.
To improve the sensitivity, logos were computed keeping into account
reverse-complement sequences and the entropy of mixing of the sets of binding
sites of the two TFs under exam (see Supplementary Note S4).

\vspace{1cm}

\begin{acknowledgments}
We thank M.~Lercher for generously providing and illustrating data
from ref.~\cite{PPL05}, H.~Salgado for help with the regulonDB
dataset, B.~Vischioni, U.~Alon, F.~Poelwijk, P.~ten~Wolde, J.~Widom,
M.~Vergassola and F.~Kepes for stimulating discussions.
\end{acknowledgments}


\vskip 0.5cm

\end{document}